\documentstyle[11pt,paspconf,psfig]{article}

\begin{document}

\title{Sgr A* and its siblings in nearby galaxies}

\author{Heino Falcke}
\affil{Department of Astronomy, University of Maryland, College Park,
MD 20742-2421, USA (hfalcke@astro.umd.edu)}




\begin{abstract}
We have proposed previously that Sgr A* is simply a scaled down AGN
with a black hole, an accretion disk and a radio jet operating at a
very low power. It appears as if M81* -- the nuclear source in the
nearby galaxy M81 -- is an ideal laboratory to study a Sgr~A*-like
source at a higher power level. The jet/disk model can explain M81* in
great detail with no basic changes in the model parameters other than
the accretion rate. Radio cores in other LINERs may be explained by
the same model and they appear to be low-power counterparts to
radio-loud quasar cores.  For Sgr A*, models without a supermassive
black hole are facing difficulties -- some of which are discussed
here, but a persistent puzzle in any scenario are the non-detections
and low flux limits for Sgr A* from IR to x-rays. Especially the IR
limits are a threat to accretion models. I discuss whether a thin
molecular disk (as seen in NGC 4258) around Sgr A* could intercept
infalling material before it reaches the black hole.

\end{abstract}




\section{Introduction}
The Galactic Center (GC) is a unique place in our galaxy, however, it is
not necessarily a unique place in the universe. For this reason the GC
has often been used as an analogy for other galaxies, and the GC can
help us to understand what we do not understand in more distant
places. But for some aspects of the GC itself, the GC is not
necessarily the best place to look for answers.

The latter is especially true for the central compact radio nucleus
Sgr~A*. While its basic radio properties are known for quite a while,
the search for counterparts in other wavelength regimes has been
largely unsuccessful. This is mainly due to the intrinsic weakness of
the Galactic Center and the obscuration in the Galactic plane. Those
difficulties have in part driven the developments of many new
instruments and techniques -- the GC is often among the first objects
to be observed with new cameras. And once in a while this has led to a
detection of Sgr A* at frequencies inaccessible to radio astronomers
(e.g. IR, NIR, X-rays, 511 keV line, etc.), but whenever the next
generation of instruments provided higher sensitivity and resolution,
it was shown that this emission was due to stellar objects and not due
to the suspected supermassive black hole in the very center. This
means that any successful model for Sgr A* has not only to be
self-consistent but must also be stable against the annual variations
in detections of Sgr A* which are a function of wavelength and spatial
resolution. Fortunately, at least the evidence from dynamical
estimates for the presence of a dark mass of
$M_\bullet=2\cdot10^6M_\odot$ in the center of the Galaxy has become
more and more convincing (Genzel 1996, Rieke \& Rieke 1996, Haller et
al. 1996) in recent years and now seems well established.

The presence of such a large concentration of dark mass in the very
nucleus of the Galaxy places the GC in line with many other galaxies
where similar or even much higher dark mass concentrations have been
found (Kormendy \& Richstone 1995). Another similarity to other
Galactic Nuclei is the presence of a compact flat spectrum radio core
which is found in all radio loud active galaxies, as well as in many
other active galaxies like radio quiet quasars, Seyferts, LINERs, in
many elliptical galaxies, and also in some spiral galaxies. In this
respect is the Galactic Center fairly typical and therefore should not be
considered as an isolated case. In this paper I will therefore not
only discuss possible explanations for Sgr A* and their difficulties,
but also apply our Sgr A* jet/disk model to other weakly active
galaxies, specifically to the nucleus of M81.

\section{Modelling Sgr A*}

\subsection{What we see...}
Sgr A* is constrained by what we see and also by what we do not see.
The size and spectrum of the radio core are the primary input data to
all models, but even though this is one of the few things we see, both
are controversial in some details. The size of Sgr A* is dominated by
scatter broadening at frequencies at least up to 22 GHz and the
smallest sizes reported so far are 1.7-2.8 AU (Doeleman et al. 1996,
Rogers et al. 1994, Krichbaum et al. 1994) at $\lambda$3mm.  The
overall spectrum of Sgr A* is inverted. While Duschl \& Lesch (1994)
claim a spectral shape of $\nu^{1/3}$ from cm to submm wavelengths,
Mark Morris presented during the conference a spectrum of Sgr A* which
was taken within 2 weeks by the CSO/JCMT/OVRO collaboration and the
VLA, indicating that the cm part may be fitted by a single powerlaw,
while the submm part shows a submm-excess. The possibility of a
submm-excess has been around for quite a while (Zylka et al. 1992) but
due to the variability of Sgr A*, which is well established in the
radio regime, it was never unambigiously proven. A clear experiment to
demonstrate this, would be truly simultaneous cm/mm/submm observations
of the GC, and preparations for such a campaign are on the way.

The importance of the submm-excess is that, if it exists, it implies
synchrotron self-absorption at frequencies around 100 GHz, and as
shown e.g. in Falcke (1996a) this requires an ultra compact region of
$\sim$ 0.1 AU. Given the mass of $2\cdot10^6M_\odot$ this size
translates into a region which is only 2-3 Schwarzschildradii in
diameter. A proper determination of the mm/submm spectrum of Sgr A*
could fix this number to a relatively high degree. Consequently, if
there really is a black hole, a future global submm VLBI experiment
would be able to probe a region which is strongly affected by General
Relativistic effects. Light bending and asymmetries due
to the Kerr Metric could in principle be directly imaged. Even though
the technical realization of such an experiment may be decades ahead,
the principal feasibility warrants a lot of excitement and motivation
for future work in the GC -- we do not know many other places, if any
at all, where such an experiment might ever become possible.

Observations of the better known cm part of the spectrum reveal kinks
(Wright \& Backer 1993) and strongly varying spectral indices (Zhao et
al. 1996, in prep.) which requires synchrotron self-absorption also at
lower frequencies and argues for a stratified medium in Sgr A* rather
than a single component model. In fact, this is what most models for
Sgr A* actually imply; and because the observed quantities of Sgr A*
are so few, but very constraining, the published Sgr A* models do not
really differ in their underlying physical processes.  The basic
ingredients of these models are a black hole and an accretion flow, a
certain conversion factor between the accretion power and the
non-thermal emission, and various equipartition arguments. In the
Bondi-Hoyle accretion model of Melia (1992 \& 1994) and in the
advection dominated disk model by Narayan et al. (1995) the radiative
efficiencies are fairly low and the emission mechanism for Sgr A* is
cyclotron/synchrotron emission, while in the jet/disk-symbiosis model
by Falcke et al. (1993a) the radiative efficiency is fairly high and
the emission mechanism is pure synchrotron radiation. In the former
cases one has pure inflow, in the latter case one has inflow and
outflow. Duschl \& Lesch (1994) proposed a stationary, homogenous blob
of synchrotron radiating monoenergetic electrons, which they
qualitatively link to an accretion disk.

\subsection{ ... and what we do not see}
As mentioned above, the basic theme for all models so far is accretion
onto a supermassive black hole and it is very difficult to avoid
strong thermal radiation from the accretion flow. Not long ago it was
thought that a substantial fraction of the central luminosity of
$10^7L_{\odot}$ in the GC is produced by this process. Alas, it is now
apparent that the stars we see are enough to produce the bulk of this
luminosity and the heating of the ambient medium and the CND (e.g.
see Latvakoski et al. 1996). The hope that some thermal emission from
Sgr A* had been discovered at least in the NIR (Eckart et al. 1992)
also faded recently with improved astrometry and resolution (Genzel
1996).  Hence, there is currently {\it no direct evidence for any
thermal emission from Sgr A*}. That may not mean much, because there
are also many luminous O/B stars in the GC which we cannot see, just
because of obscuration. Nevertheless, the constraints on accretion
disks are severe. The current NIR limits and mass estimates of the
black hole require accretion rates to be $<10^{-8}M_\odot/$yr for a
standard accretion disk (see Falcke et al. 1993a). On the other hand
Melia (e.g. 1992) has argued that a supermassive black hole in the GC
should accrete of the order $10^{-4}M_\odot$/yr from stellar winds. He
suggested spherical accretion and the formation of only a small
transient disk, Narayan et al. (1995) have proposed that the accretion
disk is advection dominated and therefore more than $99.9\%$ of the
energy is advected and never radiated. With the ever decreasing limits
on the dereddened Sgr A* flux of less than 20mJy (inferred from Eckart
et al. 1995), however, the ``inefficiency levels'' for the latter two
models become uncomfortably high, while the accretion rate for
standard accretion disk models also become uncomfortably low (even
though I have to admit that ``comfort'' is not really a well defined
physical quantity).

\subsection{Spherical accretion and fossil disk} Recently we have
proposed another alternative (Falcke \& Melia 1996) which builds up on
the suggestion by Falcke \& Heinrich (1994) that Sgr A* might well be
surrounded by a fossil accretion disk --- a remain of past
activity. Such a fossil disk (or ring) may be very optically thick,
very stable over a long time and could in principle capture infalling
matter at a large radius, possibly without producing the amount of
luminosity usually expected if the matter were to fall into the very
center. The disk would not be disrupted by the infalling wind because
it could still be very massive. The dynamical timescale of the fossil
disk is much shorter than the time scale for the infall, which is
given by the ratio $\dot \Sigma_{\rm wind}/\Sigma_{\rm disk}$ between
the mass deposition rate per unit area and the surface density of the
disk.

To study this process in more detail, we have modified the standard
accretion disk equations to allow for matter and angular momentum
deposition, and have calculated the time evolution and spectrum of
such an accretion disk. In a second step we have coupled this
accretion disk model with 3D hydrodynamical calculations of a
Bondi-Hoyle flow (Coker et al. 1996).  This allowed us to test which
scenario might be compatible with current observations. The boundary
condition was that the high mass inflow that had been intercepted at a
large radius should not propagate into the center within a couple
million years --- the presumed age of the high mass-loss star --- and
the luminosity should not exceed the current IR/NIR and total
luminosity limits.

The first result is that it is {\it not} possible to hide any strong
inflow with zero angular momentum. While indeed a lot of the wind is
captured by the fossil disk at larger radii, the remaining part of the
wind is still too large and would produce an enormous amount of
luminosity. Another problem with a zero angular momentum wind is that
even if it is absorbed at a large radius, the kinetic energy
dissipated in the impact will produce strong emission.

Therefore, one has to invoke a non-zero angular momentum wind, which
circularizes at a large radius. We find that a minimum radius for the
circularization need to be of the order $R_{\rm circ}\ga10^{16}$ cm
(i.e. 0.1''), but could of course be larger. The viscous timescale for
changes in such a structure could be as long as several 10-100 Million
years. The resulting spectrum differs substantially from a normal
accretion disk spectrum and shows a strong peak in the IR. A
complication for the modelling is that the accretion radius of the
Bondi-Hoyle flow is $10^{17}$cm and therefore any fossil structure of
the size discussed here might already start to influence the whole
Bondi-Hoyle structure and the bow shock. Another problem is that the
Bondi-Hoyle spherical accretion solution changes if one adds angular
momentum to the flow, the wind which is finally captured does not
retain the same specific angular momentum it had before it encountered
the black hole. Therefore one can not easily translate an uneven, or
rotating source distribution into an angular momentum of the infalling
wind.

From Maser observations of NGC 4258 (Miyoshi et al. 1995) we know that
molecular disks exists on such small scales. But the structure we have
discussed for Sgr A* so far is pure fantasy and was born out of the
need to explain what we do not see, and is not based on any positive
detection. Nevertheless, if present a fossil disk should of course
have observational consequences, especially in the IR. Stolovy et
al. (1996) have announced the detection of a source at $8.7\mu$m with
dereddened flux of $\sim100$ mJy at the position of Sgr A*, but at
the current resolution a direct association with Sgr A* in this
crowded and tricky field is by no means certain. Such a flux would
correspond to a black-body disk with radius $10^{15}$ cm, inclined by
$80^\circ$ at a temperature of 350$^\circ$K (not quite room
temperature but close). A disk with $R\sim10^{16}$cm at that
temperature would already produce too much flux. This shows that the
current limits for such a configuration are already very tight.

\subsection{Are there alternatives?}
With the problems currently troubling accretion models one is tempted
to ask whether there are alternative models for Sgr A*.  Is there a
life without a black hole? Can we replace the black hole with an
ultradense cluster of stellar remnants? Well, it wouldn't be much fun in the
first place.  Secondly, while one may find alternative solutions by
just considering the emission properties it becomes very difficult if
one takes the whole context into account. There is for example the
observation by Backer (1996) that Sgr A* does not move w.r.t. to the
Galactic Center, while any low mass object should (as we now see
directly in the stars). The total mass of Sgr A* therefore needs to be
at least several 100 $M_\odot$ and this mass can obviously not be in
the synchrotron radiating gas, so that we need an anchor of at least
several hundred, possibly thousand, stellar remnants. On the other hand
we need at least $10^{3.5}L_\odot$ for the synchrotron radiation
alone. If we assume that this energy comes from accretion onto this
hypothetical central cluster we have to have at least an accretion
rate of $10^{-4}M_\odot$/yr. The minimum size scale for Sgr A* is
$10^{12}$cm (see Falcke 1996a) and the radiative efficiency for a 1000
$M_\odot$ object at this scale is only $6\cdot10^{-5}$, because we are not very deep
in the potential well.  However, if the accretion continues onto the
stellar remnants, they would inevitably turn into strong x-ray
emitters. In fact, a fraction of only $10^{-6}$ of this accretion rate
would be enough to violate all current x-ray limits (see Koyama et
al. 1996, Maeda et al. 1996).

Can we then power Sgr A* without accretion, e.g. if the plasma is
heated in some mysterious way by the kinetic or potential energy of
the supposed stellar remnants? The problem here, as well as for the
accretion scenario above, is that the pressure one derives for the
synchrotron radiating gas in Sgr A* (especially in the submm where we
have $n\sim10^4{\rm cm}^{-3}$, $B\sim10$G, $r\sim10^{13}$ cm, see
e.g. Falcke 1996a) would require a central mass of $10^{8-9}M_\odot$
to keep it in the center --- otherwise it would literally be blown away
within seconds, just like in the jet-model (see also an earlier
discussion in Reynolds \& McKee 1980).  A cluster of stellar remnants
would never have enough potential energy to keep the gas in the
center. Consequently, even if there are a bunch of stellar remnants
throughout the central star cluster as suggested by Haller et
al. (1996), Rieke \& Rieke (1996), and Saha et al. (1996) it appears very
unlikely that they have anything to do with Sgr A* itself.

\section{The siblings}
\subsection{M81*}
We are strongly limited in our modelling of Sgr A* by two important
effects: scatter broadening and obscuration. Thus we know neither
the intrinsic shape and size of Sgr A*, nor its optical/UV
properties. However, as mentioned in the beginning, any model for Sgr
A* should be invariant to translation by at least a few
Mpc. Therefore, it seems as if the best place to learn more about Sgr
A* is the nucleus of M81 (see Falcke 1996b). This is a spiral galaxy,
classified as a LINER, where we are not strongly affected by
obscuration. In the nucleus we find a compact flat-spectrum radio core
(which we call M81* in analogy to Sgr A* and M31*) with a size
of 550 AU at 22 GHz and an inverted spectrum
($F_\nu\propto\nu^{0.2\pm0.2}$, see Reuter \& Lesch (1996) and
references therein). Unlike Sgr A*, this core is resolved with VLBI at
various frequencies and shows a size proportional to
$\nu^{-0.8\pm0.05}$ (Bietenholz et al. 1996, and references therein),
hence it is not scatter broadened. Moreover, the core is elongated and
one finds structure with the VLA at a much larger scale in a similar
direction. The most likely explanation for this observation is the
presence of a jet.

The bolometric luminosity of the M81 nucleus has been estimated to be
of the order $10^{41.5}$ erg/sec (Ho et al. 1996) and Bower et
al. (1996) recently discovered broad double-peaked H$\alpha$ emission
from M81, which is either due to an accretion disk or a
bi-polar outflow.

With this information it was of course tempting to apply the jet/disk
symbiosis model we developed initially for Sgr A* (Falcke et
al. 1993b) to M81*, at least here it is much easier to argue that a
mini-AGN with jet and accretion disk really is present.  Especially
the detailed VLBI informations allow a more detailed test of the
model.

The first important point is the frequency dependence of the size,
with a size index $m=-0.8\pm0.05$ ($r\propto\nu^m$) and an inverted
spectrum. The frequency dependence of the size was one of the basic
predictions of the jet model, while in homogenous, optically thin
models (e.g. Duschl \& Lesch 1994) a constant size is expected -- this
reflects the main differences between homogenous and inhomogenous
(i.e. with gradient in magnetic field) models.

However, the extremely simplified jet emission model also does not fit
perfectly, as it predicts a flat ($\alpha=0$) rather than an inverted
spectrum, and the predicted size index is $m=-1$, thus slightly
steeper than observed in M81*. It is of course fairly easy to modify
the jet model to fit those values, e.g. by imposing a certain
non-conical jet shape (as it is frequently done for quasar
cores). Such a non-conical shape would imply external confinement or
acceleration of the jet. On the other hand those models usually lack a
physical justification for the acceleration or collimation (especially
with the high internal pressures involved) and it makes one feel
uncomfortable to just add another arbitrarily chosen input parameter
for each new observed quantity. Fortunately, it turned out that there
is a slight inconsistency in the canonical Blandford \& K\"onigl
(1979) jet model used previously (e.g. Falcke \& Biermann 1995), where
one usually neglects the dynamical effects of the pressure gradient on
the velocity field of the jet flow. If calculated self-consistently
this pressure gradient will indeed lead to a slight acceleration of
the jet. In terms of the velocity structure this is a weak effect,
however, if one starts with a fully relativistic gas (i.e. sound
speeds of the order $0.6c$ -- something necessary to escape from the
inner parts of a black hole) it is just enough to make the jet mildly
relativistic ($\gamma_{\rm j}\simeq2-3$). Due to the boosting effect,
the emission at lower frequencies, coming from more distant regions,
will be Doppler-{\it dimmed} w.r.t.~the higher frequencies for most
aspect angles and thus yield an inverted spectrum and a flatter size
index.

For the given luminosity of M81* and the jet-power/disk-luminosity
ratio we found for quasars (see Falcke et al. 1995), the whole Sgr A*
jet/disk symbiosis model can then be boiled down to a two parameter
model, where we need only the electron Lorentz factor $\gamma_{\rm e}$
and the inclination angle $i$ as an input parameter, which on top of
that, are both fairly well constrained.

And in fact, for $\gamma_{\rm e}=220$ and $i\simeq30-40^\circ$ the
model predicts the observed size (550 AU at 22 GHz), flux (110 mJy at
22 GHz), spectral index ($\alpha=0.17$), and size index ($m=-0.9$) for
M81* reasonably well. Sgr A* can be explained by the same model for an
assumed $L_{\rm disk}\sim10^{39}$erg/sec with $i\sim60^\circ-70^\circ$
and $\gamma_{\rm e}=140$, the predicted average spectral index is
$\alpha=0.23$ and $m=-0.9$ --- the size of the major axis should be
around 6 AU at 7mm.

\begin{figure}
\centerline{\psfig{figure=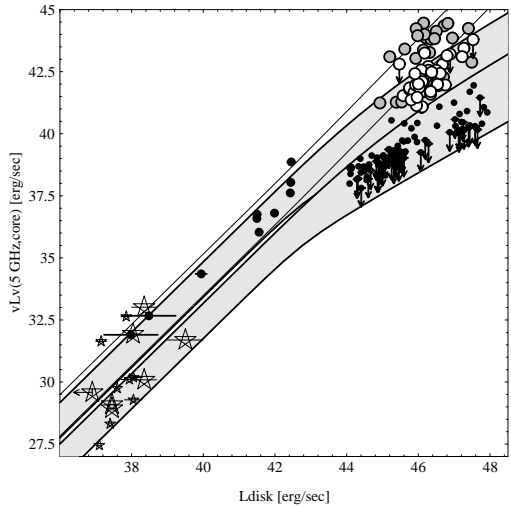,width=0.5\textwidth,bbllx=2.7cm,bblly=6cm,bburx=19.1cm,bbury=22.1cm}}
\caption{Radio core vs.~bolometric nuclear luminosity for a variety of
known and putative jet/disk systems (from Falcke \& Biermann 1996a)
and the predicted distributions from the jet/disk symbiosis model. The
stars in the lower left are galactic jet sources and x-ray binaries,
circles and dots in the upper right are radio loud and radio quiet
quasars. Black dots below $L_{\rm disk}=10^{44}$ erg/sec are LINERs
and Sgr A* and M31* -- it appears as if LINERs could be the missing
link between highly active radio loud quasars and almost inactive, yet
radio-luminous, nuclei like Sgr A*.}
\end{figure}

\subsection{The rest of the family}
It is interesting to note that we had to use a {\it radio-loud} model
(defined by the radio/$L_{\rm disk}$ ratio) to explain M81* (the same
is true for Sgr A* and possibly M31*). Could it be that the cores of
radio loud quasars have their low-luminosity counterparts in LINERs
and other weakly active galaxies?  For this reason we have begun to
revisit the radio properties of some nearby galaxies with signs of
nuclear activity; quite a few have compact flat-spectrum radio cores
similar to Sgr A* and M81*, e.g. like the Sombrero galaxy
(M104). Unfortunately, due to the low level of activity, the
determination of a bolometric or ``disk''-luminosity for the nuclei
can be very difficult.  Bearing that in mind, we have plotted the
radio core fluxes of a small sample of prominent galaxies versus what
we estimate to be their disk luminosity in Fig.1. Those results are of
course very preliminary and need further refinement, nevertheless, it
is quite interesting that the cores of those LINER nuclei all seem to
fall on the radio-loud branch of the jet/disk symbiosis model, and
some of them, like NGC 1097, do indeed have well known jets.  This
could mean that basically all those radio cores in LINERs are the
bases of radio jets and they could be the missing link between Sgr A*
and radio loud quasars. Further study of those radio cores in the
Galactic Center and elsewhere might therefore not only reveal
something about the true nature of Sgr A*, but also help us to
understand the radio-loud/radio-quiet dichotomy in quasars.

\section{Conclusion}
All the models proposed for Sgr A* have a certain appeal. The jet/disk
model -- with and without monoenergetic electrons -- offers a scope
that goes far beyond the GC and has survived a series of critical
tests in a variety of very different source classes with compact flat
spectrum cores, including Sgr A*. Advection-dominated and fossil disks
may help to explain why the optical luminosity of Sgr A* is so low,
and Bondi-Hoyle accretion is a process that seems to be unavoidable at
a certain level. Applying all these concepts to the nuclei of nearby
galaxies may help us to sort out which process dominates in which
regime. Until then we perhaps could agree on a
``theorists-for-galactic-peace-model'' for Sgr A*: a jet of
monoenergetic electrons, produced by an advection dominated disk
coming from a fossil ring which is fed by Bondi-Hoyle accretion.

\acknowledgments This work was supported by NASA under grants
NAGW-3268 and NAG8-1027. I am grateful for many discussions during and
after the conference.


%
%

%

\end{document}